Instruction: There are five figures. The postscripts files are
included at the end of the file (after \end{document}). The figures
are included in the text using \input psfig.

\documentstyle[twoside,12pt]{article}

\catcode`\@=11
\long\def\@makefntext#1{ 
\protect\noindent \hbox to 3.2pt {\hskip-.9pt
$^{{\eightrm\@thefnmark}}$\hfil}#1\hfill} 

\def\thefootnote{\fnsymbol{footnote}}
 \def\@makefnmark{\hbox to 0pt{$^{\@thefnmark}$\hss}}  

\def\ps@myheadings{\let\@mkboth\@gobbletwo
\def\@oddhead{\hbox{} 
\rightmark\hfil\eightrm\thepage}
\def\@oddfoot{}\def\@evenhead{\eightrm\thepage\hfil 
\leftmark\hbox{}}\def\@evenfoot{}
\def\sectionmark##1{}\def\subsectionmark##1{}}

\oddsidemargin=\evensidemargin
\renewcommand{\thefootnote}{\fnsymbol{footnote}}

\newcounter{sectionc}\newcounter{subsectionc}\newcounter{subsubsectionc}
\renewcommand{\section}[1] {\vspace{12pt}\addtocounter{sectionc}{1}
\setcounter{subsectionc}{0}\setcounter{subsubsectionc}{0}\noindent
	{\bf\thesectionc. #1}\par\vspace{5pt}}
\renewcommand{\subsection}[1] {\vspace{12pt}\addtocounter{subsectionc}{1}
	\setcounter{subsubsectionc}{0}\noindent
	{\bf\thesectionc.\thesubsectionc. {\kern1pt \bf\it #1}}\par\vspace{5pt}}
\renewcommand{\subsubsection}[1] {\vspace{12pt}\addtocounter{subsubsectionc}{1}
	\noindent{\thesectionc.\thesubsectionc.\thesubsubsectionc.
	{\kern1pt \it #1}}\par\vspace{5pt}}
\newcommand{\nonumsection}[1] {\vspace{12pt}\noindent{\bf #1}
	\par\vspace{5pt}}

\newcommand{\textlineskip}{\baselineskip=14pt}
\newcommand{\smalllineskip}{\baselineskip=12pt}

\def\eightcirc{
\begin{picture}(0,0)
\put(4.4,1.8){\circle{6.5}}
\end{picture}}
\def\eightcopyright{\eightcirc\kern2.7pt\hbox{\eightrm c}}

\renewenvironment{thebibliography}[1]			
	{
	 \begin{list}{\arabic{enumi}.}			
	{\usecounter{enumi}\setlength{\parsep}{0pt}
	 \setlength{\leftmargin 17pt}{\rightmargin 0pt}	
	 \setlength{\itemsep}{0pt} \settowidth		
	{\labelwidth}{#1.}\sloppy}}{\end{list}}	

\newcounter{itemlistc}
\newcounter{romanlistc}
\newcounter{alphlistc}
\newcounter{arabiclistc}

\newcommand{\fcaption}[1]{
        \addtocounter{figure}{1}
         {{\tenrm Fig.~\thefigure . #1} }\hfil\break }

\newcommand{\tcaption}[1]{                      
        \addtocounter{table}{1}
         {{\tenrm\offinterlineskip Table~\thetable . #1} }\hfil\break }

\def\pmb#1{\setbox0=\hbox{#1}
	\kern-.025em\copy0\kern-\wd0
	\kern.05em\copy0\kern-\wd0
	\kern-.025em\raise.0433em\box0}

\def\fnt#1#2{\footnotetext{\kern-.3em
	{$^{\mbox{\scriptsize #1}}$}{#2}}}

\def\fpage#1{\begingroup
\voffset=.3in
\thispagestyle{empty}\begin{table}[b]\centerline{\footnotesize #1}
	\end{table}\endgroup}

\def\go{\gamma_1}
\def\gw{\gamma_2}
\def\gt{\gamma_3}
\def\wo{\omega}

\def\alp{\alpha^\prime}

\def\bep{\beta^\prime}
\def\bpp{\beta^{\prime\prime}}
\def\app{\alpha^{\prime\prime}}
\def\gm{\gamma}

\def\gm{\gamma}

\def\dag{\dagger}

 1
 1
 1

\font\eightrm=cmr8

\def\qed{\hbox{${\vcenter{\vbox{                          
   \hrule height 0.4pt\hbox{\vrule width 0.4pt height 6pt
   \kern5pt\vrule width 0.4pt}\hrule height 0.4pt}}}$}}

\input psfig

\begin{document}
\normalsize\textlineskip
{\thispagestyle{empty}
\setcounter{page}{1}

\renewcommand{\thefootnote}{\fnsymbol{footnote}} 
\def\bsc{{\sc a\kern-6.4pt\sc a\kern-6.4pt\sc a}}
\def\bflatex{\bf L\kern-.30em\raise.3ex\hbox{\bsc}\kern-.14em
T\kern-.1667em\lower.7ex\hbox{E}\kern-.125em X}

\fpage{1}
\centerline{\bf VECTOR MESONS IN THE COLLECTIVE APPROACH TO SU(3)
$^\dagger $}
\vspace{0.37truein}
\centerline{\footnotesize HERBERT WEIGEL}
\vspace*{0.015truein}
\centerline{\footnotesize\it Institut for Theoretical Physics,
T\"ubingen University}
\baselineskip=12pt
\centerline{\footnotesize\it D-7400 T\"ubingen, FR Germany}

\vspace*{0.21truein}

\vspace*{-3pt}\textlineskip
\section{Introduction and Motivation}

Assuming the point of view that Quantum Chromo Dynamics (QCD) is
well approximated by an effective meson Lagrangian$^{1,2}$ with baryons
emerging as soliton solutions$^{3,4}$ we shall investigate such a
Lagrangian for the case of three flavors.
The exploration of SU(3) models is very interesting since it
provides extensive information about the low lying $\frac{1}{2}^+$ and
$\frac{3}{2}^+$ baryons. Of course, the first results extracted
from such a model concern the baryon mass spectrum, however, it
is possible to go one step further and obtain predictions for static
properties like electromagnetic form factors or matrix elements
of the axial current which are measured in semileptonic hyperon decays.
In view of recent or upcoming experiments it is very challenging
to consider the structure of the proton in a three flavor model. At
this point the famous EMC measurement$^5$ of scattering polarized muons
off polarized protons which led to the ``proton spin puzzle" and
experiments on electron-hadron scattering via neutral currents$^6$ are to
be mentioned. SU(3) models furthermore may shed some light on the
strangeness content of the proton which appears to be an important
ingredient for the determination$^7$ of the pion-nucleon $\Sigma$ term,
$\Sigma_{\pi N}$ as well as for parity violation processes$^8$.

\vfill
\noindent
$^\dagger $ Invited talk presented at the international workshop
on ``Baryons as Skyrme Solitons", Siegen, Sept. 92.
\eject

The first step in constructing an effective meson action
naturally consists of employing the non-linear $\sigma-$model which allows
for solitonic solutions once stabilizing terms are added.
Most commonly Skyrme's$^3$ antisymmetric fourth order term is
employed for this purpose but also a six-order term has frequently
been used.$^9$ Since these additional terms are of higher orders in
the derivative operator and are only restricted by demanding
them to be chirally symmetric they introduce arbitrary parameters which
cannot be determined from mesonic observables. However, stability
of the soliton may as well be gained by the incorporation of
vector meson fields like $\rho$ and $\omega$.$^{10,11}$ The stability is then
provided by the ``anomalous" terms in the action which contain the
Levi-Cevita tensor $\epsilon_{\mu\nu\rho\sigma}$. In the meson
sector these terms describe vertices like $\omega\rightarrow\pi\pi\pi$
or $\omega\rightarrow\rho\pi$. The experimentally determined decay
widths for these processes fix the associated parameters. An
additional coupling constant appearing in the non-anomalous sector
may be determined from the decay $\rho\rightarrow\pi\pi$.
Furthermore it is important to note that for the description of baryons
in the SU(2) model vector mesons have been proven to be necessary
ingredients to improve the predictions for static properties of the
nucleon$^{12}$ ({\it e.g.} electromagnetic form factors), the
neutron proton mass difference$^{13}$ and the ``high" energy behavior
of the phase shifts in $\pi N$ scattering.$^{14,15}$

Unfortunately the extension to SU(3) cannot be done straightforwardly
due to the presence of flavor symmetry breaking. In models consisting
of pseudoscalar mesons only, two (on first sight quite different)
treatments of SU(3) symmetry breaking in the baryon sector have
been relatively successful. In the bound state (BS) approach
collective coordinates are only introduced for the rotations into
``non-strange" directions, the real zero modes.$^{16}$ Canonical quantization
of these coordinates yields states with good quantum numbers for
spin and isospin. A kaonic bound state in the background field of the
SU(2) soliton is constructed and each occupation of this bound state
lowers the hypercharge by one unit. On the other hand, in the Yabu and Ando
(YA) approach to symmetry breaking SU(3) is considered approximative
and collective coordinates are therefore introduced for the
complete SU(3) rotation.$^{17}$ Employing an ``Euler angle" parametrization of
these rotations the collective Hamiltonian which contains all
relevant information about flavor symmetry breaking may be diagonalized
exactly. The YA treatment furthermore allows to easily take the
zero symmetry breaking limit and thus it enables one to compare with
ordinary SU(3) baryon phenomenology.$^{18}$ This comparison shows that
the ${1\over2}^+\ ({3\over2}^+)$ baryons are no longer pure octet
(decouplet) states but contain significant admixtures of higher
dimensional representations like ${\overline{10}}$ or $27$
$(27,35,{\overline{35}})$. A detailed study of the YA approach in
models with pseudoscalar mesons only$^{19,20}$ has shown that symmetry
breaking effects are large for mass differences and matrix elements
of strangeness conserving operators but negligible for those of
strangeness changing operators, needed e.g. in the Cabibbo model
of hyperon decays. In view of the success of the YA treatment in
the pseudoscalar model we shall generalize this approach to
models containing vector mesons as well.

Gauging the vector meson Lagrangian with external fields allows
to read off the corresponding currents. For the coupling of vector
mesons to photons from the non-anomalous terms this approximately
yields the phenomenologically successful concept of vector meson
dominance (VMD).$^{21}$ It is therefore very interesting to
investigate electromagnetic properties of strange and non-strange
baryons since the predictions for magnetic moments and electromagnetic
radii are expected to improve due to VMD.$^{12,22}$ In pure
pseudoscalar models the magnitudes of these observables commonly come
out too small.$^{23}$ Furthermore the vector meson model appears to give
more reliable information on the axial singlet current matrix
element$^{24,25}$ as well as the two flavor contribution to the neutron
proton mass difference$^{13}$ which both vanish identically in
pseudoscalar models.$^{13,26}$

This talk is organized as follows. In section 2) we shall construct the
mesonic action as well as the soliton solution in the baryon sector. The
baryon mass spectrum is explored in section 3). Section 4) contains the
the discussion of the static properties of the low-lying
$\frac{1}{2}^+$ and $\frac{3}{2}^+$ baryons. A few conclusive remarks
are found in section 5).

\textheight=8.5truein
\setcounter{footnote}{0}
\renewcommand{\thefootnote}{\alph{footnote}}

\section{The Model}

\subsection{The mesonic part}

The mesonic action, $\Gamma$, which we shall adopt here, was first presented
in ref.$^{27}$ in its manifest chirally invariant form. The authors
employed the non-linear realization $U={\rm exp}(i\sqrt2\phi/f_\pi)$ of
the pseudoscalar nonet field $\phi$ as well as left and right
transforming vector mesons fields $A_\mu^L$ and $A_\mu^R$, respectively.
The axial vector meson was eliminated by imposing the non-linear constraint:
\begin{eqnarray}
A_\mu^R=U^\dagger A_\mu^LU+{i\over g}U^\dagger\partial_\mu U,
\end{eqnarray}
$g$ being a coupling constant to be determined later. The constraint (1)
may be regarded as the defining equation for the vector meson nonet field
$\rho_\mu$ by writing:
\begin{eqnarray}
A_\mu^L &=\xi\rho_\mu\xi^\dagger+{i\over g}
\xi\partial_\mu\xi^\dagger\cr
A_\mu^R &=\xi^\dagger\rho_\mu\xi+{i\over g}
\xi^\dagger\partial_\mu\xi,
\end{eqnarray}
wherein $\xi=U^{1\over2}$. Defining furthermore, for convenience:
\begin{eqnarray}
&p_\mu&=\partial_\mu \xi\xi^\dagger +\xi^\dagger \partial_\mu\xi ,
  \quad v_\mu= \xi^\dagger \partial_\mu\xi-\partial_\mu \xi\xi^\dagger  \cr
&R_\mu&=\rho_\mu-{i\over{2g}}v_\mu,\quad F_{\mu\nu}(\rho)=\partial_\mu\rho_\nu
 -\partial_\nu\rho_\mu-ig[\rho_\mu,\rho_\nu],
\end{eqnarray}
the quantities $v_\mu$ and $p_\mu$ obviously transform as vector and
pseudovector fields.

We may now list the mesonic action $\Gamma$ by splitting it up into four
pieces:
\begin{eqnarray}
\Gamma=\int d^4x\big({\cal L}_S+{\cal L}_{SB}+{\cal L}_G\big)
+\Gamma_{an}.
\end{eqnarray}
The symmetric, non-anomalous part of the Lagrangian is given by:
\begin{eqnarray}
{\cal L}_S=-{{f_\pi^2}\over4}Tr(p_{\mu}p^{\mu})
-{1\over 2}Tr[F_{\mu\nu}(\rho )F^{\mu\nu}(\rho)]
+m_V^2 Tr(R_\mu R^\mu ),
\end{eqnarray}
while the anomalous part of the action is most clearly arranged by using the
notation of one-forms ($p=p_\mu dx^\mu$, etc.):
\begin{eqnarray}
\Gamma_{an}&&={{iN_c}\over{240\pi^2}}\int Tr(p^5)\cr
&&\quad +\int Tr[{1\over 6}(\go +{3\over 2}\gw ) Rp^3
-{{ig}\over 4} \gw F(\rho )
(pR-Rp)-g^2 (\gw+2\gt) R^3p],
\end{eqnarray}
wherein the first expression on the $RHS$ represents the Wess-Zumino term, an
integral over a five dimensional manifold whose boundary is Minkowski space.

To complete the SU(3) action we include the symmetry breaking terms as
discussed in ref.$^{13}$ which, by using the parametrization (3), acquire the
form:
\begin{eqnarray}
&{\cal L}_{SB}=Tr\Big[-2 R_\mu R^\mu (\alp T^++\app
S^+)-{i\over g}[R_\mu,p^\mu](\alp T^-+\app S^-)\cr
&+p_\mu p^\mu [(\bep-{\alp\over{2g^2}})T^+
+(\bpp-{\app\over{2g^2}})S^+]+{\delta^\prime}(T^+-2T) +
{\delta^{\prime\prime}}(S^+-2S)\Big],
\end{eqnarray}
where we defined, for convenience,
$ T^{\pm}=\xi T\xi \pm {\xi^\dag}T {\xi^\dag} ,\quad
S^{\pm}=\xi S\xi \pm {\xi^\dag} S {\xi^\dag} $
and $T=diag(1,1,0),S=diag(0,0,1)$. Under chiral $SU(3)_L\times SU(3)_R$,
${\cal L}_{SB}$ transforms according to the representation
$(3,3^\ast)+(3^\ast,3)$.

The parameters appearing in ${\cal L}_S$ and ${\cal L}_{SB}$ may be
determined by noting, {\it e.g.}, that the physical pion decay constant
$f_{\pi p}=0.093GeV$ and the vector meson masses $m_\rho=m_\omega=0.773GeV$
are related to the parameters in (5) and (7) via:
\begin{eqnarray}
f_{\pi p}^2=f_\pi^2+4({\alp\over{g^2}}-2\bep),\quad
m_\rho^2=m_V^2-4\alp.
\end{eqnarray}
Together with the condition that the symmetry breaking matrices are
proportional to the current quark mass matrix ${\cal M}=
diag(m_u,m_d,m_s)$, imposing:
\begin{eqnarray}
\alpha^\prime :\alpha^{\prime\prime}=
\beta^\prime :\beta^{\prime\prime}=
\delta^\prime :\delta^{\prime\prime}\stackrel{!}{=}
{\hat m}:m_s.
\end{eqnarray}
with ${\hat m}=\frac{1}{2}(m_u+m_d)$. At this point it is important to
mention that the quark masses do {\it not} serve as input quantities,
on the contrary, the ratio ${\hat m}:m_s$ will be a prediction.
This ratio is identical to the quantity $x$ in ref.$^{28}$. Noting
furthermore that the third term in (5) contains the
$\rho\pi\pi$ vertex, which may be used to determine $g$, one finds:$^{13}$
\begin{eqnarray}
f_\pi&\approx&0.092GeV,\quad m_V\approx0.766GeV,\quad g\approx5.57,\cr
\alp&\approx&-0.0028GeV^2,\quad  \app\approx -0.102GeV^2,\cr
\bep&\approx&-7.14\times 10^{-5}GeV^2,\quad
\bpp\approx-0.00266GeV^2,\cr
{\delta^\prime}&\approx&4.15\times 10^{-5}GeV^4,\quad
{\delta^{\prime\prime}}\approx1.55\times 10^{-3}GeV^4,
\end{eqnarray}
wherein errors due to uncertainties in the decay constants and masses
are not denoted.

In ref.$^{11}$ two of the three unknown constants $\gamma_{1,2,3}$ were
determined from purely strong interaction processes. Defining
$\tilde h=-{{2\sqrt2}\over3}\gamma_1$, $\tilde g_{VV\phi}=g\gamma_2$ and
$\kappa={{\gamma_3}\over{\gamma_2}}$ the central values $\tilde h=0.4$ and
$\tilde g_{VV\phi}=1.9$ were found. Within experimental uncertainties
(stemming from the uncertainty in the $\omega - \phi$ mixing angle) these
may vary in the range $\tilde h=-0.15,..,0.7$ and $\tilde g_{VV\phi}=1.3,..,
2.2$ subject to the constraint $\vert\tilde g_{VV\phi}-\tilde h\vert\approx
1.5$. The third parameter, $\kappa$ could not be fixed in the meson sector,
however, from the study$^{12}$ of nucleon properties in the U(2) reduction of
the model it was argued that $\kappa\approx1$. Furthermore note that the
overall signs of $\gamma_i$ cannot be fixed from the knowledge of mesonic
properties.

Finally we may add a term to mock up the $U(1)$ anomaly:$^{29}$
\begin{eqnarray}
{\cal L}_G=\frac{G^2}{12f_\pi^2m_{\eta^\prime}^2}
+\frac{i}{12}G\ {\rm ln}\Big(\frac{{\rm det}U}{{\rm det}U^\dagger}\Big).
\end{eqnarray}
The pseudoscalar glueball $G$ is related to the QCD field strength
tensor $F_{\mu\nu}$ and coupling constant $g_{QCD}$ by:
\begin{eqnarray}
G=\frac{3ig_{QCD}^2}{16\pi^2}\epsilon_{\mu\nu\rho\sigma}
{\rm tr}\big(F^{\mu\nu}F^{\rho\sigma}\big)
\end{eqnarray}
obviously there is no kinetic term for $G$ and thus it may be
eliminated via the equation of motion:
\begin{eqnarray}
G=2\sqrt3m_{\eta^\prime}^2\eta^\prime
\end{eqnarray}
providing the mass term for pseudoscalar singlet $\eta^\prime$.

\subsection{The baryon sector}

We assume the well established point of view that baryons emerge as
soliton solutions of the mesonic theory$^{2,3,4}$. In the model under
consideration this solution is found$^{10}$
by invoking the hedgehog {\it ans\"atze} in the isospin subgroup of SU(3):
\begin{eqnarray}
\xi_\pi({\bf r})={\rm exp}\big(i{\bf \hat r}\cdot
{\mbox{\boldmath $\tau$}} F(r)/2\big),\quad
\omega_0={{\omega(r)}\over{2g}},\quad
\rho_{i,a}={{G(r)}\over{gr}}\epsilon_{ija}\hat r_j.
\end{eqnarray}
Substitution of these {\it ans\"atze} yields the classical mass $M_{cl}$.
An analytic expression for $M_{cl}$ may be found in appendix A of ref.$^{30}$.
Minimization of $M_{cl}$ yields Euler-Lagrange equations which are
integrated in conjunction with boundary conditions appropriate to the
baryon number one sector. The resulting profiles $F,\omega$
and $G$ are shown in figure 1.
\begin{figure}
\centerline{\hskip -1.5cm \psfig{figure=vem.ps,height=9.0cm,width=16.0cm}}
\fcaption{The profiles of the classical soliton. $\omega$ is
measured in $GeV$. The input parameters are as in eq. 28.}
\end{figure}

The field configuration (14) obviously does not lead to states with
either good spin or isospin. In order to project on spin and SU(3)
quantum numbers the zero modes are quantized semiclassically via the
introduction of suitable time dependent collective coordinates. As
motivated in the introductory section, we consider SU(3) as an
approximative symmetry. Therefore we introduce collective coordinates for
the rotations in SU(3) and not only $SU_I(2)\times U_Y(1)$ which are
the real zero modes of the Lagrangian.

Special care is required concerning fields which vanish as long as only static
configurations are considered. These fields may, however, be excited
by the time-dependent rotation of the classical fields $F,\omega$ and $G$.
Besides introducing collective coordinates $A(t)\in SU(3)$ we
generalize the {\it ans\"atze} for the fields in the body-fixed frame
via:$^{19}$
\begin{eqnarray}
\xi({\bf r},t)=A(t)\xi_k\xi_\pi({\bf r})\xi_kA^\dagger(t)
\end{eqnarray}
The unitary (but not unimodular) matrix $\xi_k$ is most
conveniently parametrized by
\begin{eqnarray}
\xi_k=e^{iz},\quad z=\pmatrix{\eta_T &K\cr K^\dagger&\eta_S\cr}.
\end{eqnarray}
For the vector fields we write:$^{30,31}$
\begin{eqnarray}
\rho_\mu({\bf r},t)=A(t)\pmatrix{\rho^\pi_\mu+\wo_\mu &K^*_\mu \cr
{K^{*\dag}_\mu}& 0\cr}A^\dagger(t).
\end{eqnarray}
The time dependence of the collective rotations is exhibited most
clearly by defining angular velocities $\Omega_a$ (a=1,...,8):
\begin{eqnarray}
A^{\dag}{\dot A}={i\over2}{\sum^8_{a=1}}{\lambda_a}{\Omega_a}=
i\pmatrix{{\Omega_\pi}+{\Omega_\eta}&\Omega_K\cr
{\Omega_K^{\dag}}&-2{\Omega_\eta}\cr}.
\end{eqnarray}
The pseudoscalar nonet contains both, components which are excited
by the isospin rotation
\begin{eqnarray}
\eta_T=\frac{1}{4}\big(\chi+\chi_8\big){\bf \hat r}\cdot{\bf \Omega}, \quad
\eta_S=\frac{1}{4}\big(\chi-2\chi_8\big){\bf \hat r}\cdot{\bf \Omega}, \quad
\end{eqnarray}
as well as components which are excited by
the rotations into strange direction
\begin{eqnarray}
K=W(r){\bf \hat r}\cdot{\mbox{\boldmath $\tau$}}\Omega_K.
\end{eqnarray}
In view of the constraint (13) the decomposition of the $\eta$ fields
into singlet ($\chi$) and octet ($\chi_8$) seems to be most
appropriate. From the parity and isospin properties we get the most general
{\it ansatz} for the vector mesons which are induced by the isospin
rotation:$^{22}$
\begin{eqnarray}
\rho_0^\pi ={1\over {2g}}[\xi_1(r) {\bf \Omega} +
\xi_2(r)({\bf \hat r}\cdot {\bf \Omega}){\bf \hat r}]
\cdot{\mbox{\boldmath $\tau$}}
\quad \wo_i
={{\Phi(r)}\over {2g}}\epsilon_{ijk}\Omega_j\hat r_k.
\end{eqnarray}
Suitable {\it ans\"atze} for the vector meson fields which are excited
by the kaonic angular velocity $\Omega_K$ are given by$^{31}$
\begin{eqnarray}
K^*_0={{S(r)}\over\tilde g}\Omega_K,\quad
K^*_i={1\over {2\tilde g}}[iE(r)\hat r_i +
{D(r)\over r}\epsilon_{ijk}\hat r_j \tau_k]\Omega_K.
\end{eqnarray}

Substituting these field configurations into the action and expanding
up to second order in the angular velocities yields the
collective Lagrangian:
\begin{eqnarray}
L=&-&M_{cl}+{1\over 2}\alpha^2\sum_{i=1}^3 \Omega_i^2
+{1\over 2}\beta^2\sum_{\alpha=4}^7 \Omega_\alpha^2-{{\sqrt 3}\over 2}\Omega_8
-{\gm\over 2}[1-D_{88}(A)]\cr
&+&\alpha_1\sum_{i=1}^3 D_{8i}\Omega_i
+\beta_1\sum_{\alpha=4}^7 D_{8\alpha}\Omega_\alpha
\end{eqnarray}
$D_{ij}$ denote the rotation matrices ${1\over2}Tr(\lambda_iA\lambda_j
A^\dagger)$. Terms of the order (symmetry breaking)$\times$(angular
velocity)$^2$ have been neglected\footnote{Note that only the classical
fields are contained in $\alpha_1$ and $\beta_1$ in order to
preserve isospin invariance(cf. appendix C of ref.$^{19}$).}.

The non-strange moment of inertia, $\alpha^2$ is a functional of the
induced fields $\xi_1,\xi_2,$ $\Phi,\eta_1$ and $\eta_2$ with the
classical profiles (14) as source fields:
$\alpha^2[\xi_1,\xi_2,\Phi,\eta_1,\eta_2;F,\omega,G]$. The excitations
are obtained by extremizing $\alpha^2$ together with boundary
conditions which render $\alpha^2$ finite. The radial dependence of
the excitations is displayed in figure 2 for a typical set of
parameters (28).
\begin{figure}
\centerline{\hskip -1.5cm\psfig{figure=su2.ps,height=9.0cm,width=16.0cm}}
\fcaption{The non-strange excitations
$\xi_1,\xi_2,\Phi,\eta_1$ and $\eta_2$ as functions of the
radial distance $r$. The radial functions $\xi_1$ and $\xi_2$ are
dimensionless while $\Phi,\eta_1$ and $\eta_2$ are measured in
$GeV^{-1}$.}
\end{figure}
In the same manner extremization of the strange moment of inertia,
$\beta^2=\beta^2[W,S,E,D;F,\omega,G]$ provides the radial functions
$W,S,E$ and $D$ which are shown in figure 3. Details of these calculations
as well as explicit expression for $\alpha^2$ and $\beta^2$ may be
found in ref.$^{30}$.

\begin{figure}
\centerline{\hskip -1.5cm\psfig{figure=su3.ps,height=9.0cm,width=16.0cm}}
\fcaption{The strange excitations $W,S,E$ and $D$
as functions of the radial distance $r$. The radial functions
$S$ and $E$ are dimensionless while $W$ and $D$ are measured in
$GeV^{-1}$.}
\end{figure}

\vfil\eject

\section{The Baryon Mass Spectrum}

In order to obtain the baryon mass spectrum we need to derive the
collective Hamiltonian from the collective Lagrangian (23).
Canonical quantization is most conveniently carried out by
introducing SU(3) right generators, $R_a\ (a=1,..,8)$:
\begin{eqnarray}
R_a=-{{\partial L}\over{\partial\Omega_a}}=\cases{
-(\alpha^2\Omega_a+\alpha_1^0D_{8a})=-J_a,&a=1,2,3\cr
-(\beta^2\Omega_a+\beta_1^0D_{8a}),&a=4,..,7\cr
{1\over2}\sqrt3,&a=8,}
\end{eqnarray}
wherein $J_i\ (i=1,2,3)$, are the spin operators.
It is helpful to define the eigenvalues of
\begin{eqnarray}
C_2+\beta^2\gamma(1-D_{88})+\beta^2{{\alpha_1^0}\over{\alpha^2}}
\sum_{i=1}^3 D_{8i}(2R_i+\alpha_1^0D_{8i})+
\beta_1^0\sum_{\alpha=4}^7D_{8\alpha}(2R_\alpha+
\beta_1^0D_{8\alpha})
\end{eqnarray}
as $\epsilon_{SB}$. $C_2=\sum_{a=1}^8R_a^2$ denotes the quadratic
Casimir operator of SU(3). $\epsilon_{SB}$ may be obtained by
introducing an ``Euler angle" parametrization of the collective
rotation matrices $A$.$^{17}$ Then the generators $R_a$ are expressed
in terms of partial differential operators\footnote{Explicit expressions
are given in appendix B of ref.$^{20}$.}. Furthermore the operator (25) is
a second order differential operator whose explicit form depends on
the isospin multiplet under consideration. Integrating these
differential equations finally gives $\epsilon_{SB}$. The eigenfunctions
are ``distored" SU(3) $D$-functions. Of course, this
procedure represents just a generalization of the original YA treatment
who only took the symmetry breaker $\beta^2\gamma(1-D_{88})$ into
account.$^{17}$ The collective Hamiltonian:
\begin{eqnarray}
H=-\sum_{a=1}^8R_a\Omega_a-L
\end{eqnarray}
is now easily diagonalized exactly, resulting in the energy formula:
\begin{eqnarray}
E=M_{cl}+{1\over2}\big({1\over{\alpha^2}}-{1\over{\beta^2}}\big)J(J+1)
-{3\over{8\beta^2}}+{1\over{2\beta^2}}\epsilon_{SB}.
\end{eqnarray}
It is well known that this procedure gives too large absolute values
for the masses of the baryons. However, it has been shown in the
simple SU(2) model of pseudoscalars only$^{32}$ that there
are significant subtractions originating from quantum corrections
due to pion loops. Moreover, these subtructions are dominated by
the $O(N_C^0)$ contribution and are therefore identical for nucleon and
$\Delta$.$^{33}$  Thus it is suggestive and justified to only consider
mass differences.

In table 1 we list our numerical results for the mass differences
of the low-lying $\frac{1}{2}^+$ and $\frac{3}{2}^+$ baryons with
respect to the nucleon.
\begin{table}
\tcaption{The mass differences of various baryons with respect to the
nucleon in MeV compared to experimental data. The equation numbers
refer to the input parameters and PS denotes the results in the
pseudoscalar model of ref.$^{19}$.}
\centerline{\tenrm\smalllineskip
\begin{tabular}{l c c c c}\\
& eq. 10 & eq. 28 & Expt. & PS \\
\hline
$\Lambda$  & 136  & 159 & 177 & 157 \\
$\Sigma$   & 252  & 270 & 254 & 238 \\
$\Xi$      & 349  & 398 & 379 & 371 \\
$\Delta$   & 351  & 311 & 293 & 246 \\
$\Sigma^*$ & 467  & 448 & 446 & 381 \\
$\Xi^*$    & 586  & 592 & 591 & 520 \\
$\Omega$   & 681  & 718 & 733 & 661 \\
\hline\\
\end{tabular}}
\end{table}
The second column in table 1 displays the best fit to the mass spectrum
which is obtained for the input parameters:
\begin{eqnarray}
\tilde h&=&0.3,\quad \tilde g_{VV\phi}=1.8,\quad \kappa=1.2,\cr
\alp&=&-0.0028GeV^2,\quad \app= -0.100GeV^2,\cr
\bep&=&-7.14\times 10^{-5}GeV^2,\quad
 \bpp=-0.00270GeV^2,\cr
{\delta^\prime}&=&4.15\times 10^{-5}GeV^4,\quad
{\delta^{\prime\prime}}=1.55\times 10^{-3}GeV^4,
\end{eqnarray}
$f_{\pi p},m_V$ and $g$ remaining unchanged.
It should be noted that this set is compatible with the parameters
needed to describe the relevant observables in the meson sector.
For the central values of input parameters (10) the splitting between
multiplets of different spin is predicted too large. This is
due to the small non-strange moment of inertia $\alpha^2$.
Changing to parameter set (28) leads to a larger soliton. This
increases $\alpha^2$ and gives therefore a better fit to the
$N-\Delta$ splitting. The larger soliton also provides an
increased symmetry breaking parameter $\gamma$ leading to
a better description of the SU(3) mass differences. Obviously the
problem is reversed in the pseudoscalar model (denoted PS in
table 1): For a suitable $\gamma$ the non-strange moment of inertia
is too large.  Amusingly we find in the vector meson model that the
predicted overall mass difference in the spin $1\over2$ multiplet,
$M_\Xi-M_N=398MeV$ is larger than the experimental value ($379MeV$) while
for the spin $3\over2$ multiplet our result for the overall
splitting, $M_\Omega-M_\Delta=407MeV$ is smaller than the experimental
number ($440MeV$).
The fact that the model describes the (relative) mass spectrum reasonably
well provides a strong motivation to also explore static properties of
${1\over2}^+$ baryons.

\section{Static Properties of Baryons}

Most of the static properties of baryons are related to matrix
elements of Noether currents. It is therefore essential to extract
these currents from the Lagrangian. This is conveniently performed
by adding external gauge fields for local vector ($V$) and
axial vector ($A$) transformations:
$B_\mu^{(V,A)}=\sum_{a=0}^8 B_\mu^{(V,A)a}Q_a
\quad (Q_0=diag({1\over3},{1\over3},{1\over3})$ and
$Q_a={{\lambda_a}\over2}\
(a=1,..,8)$, with $\lambda_a$ being the Gell-Mann matrices)
such that the total action $\Gamma[U,A^L,B^{(V)},B^{(A)}]$ is
gauge invariant.$^{11}$ The covariant expressions for the
vector ($V_\mu^a$) and axial vector ($A_\mu^a$) currents are then
read off as the coefficient of the terms linear in the gauge fields:$^{12}$
\begin{eqnarray}
V_\mu^a={{\delta\Gamma(U,A^L,B^{(V)},B^{(A)})}
\over{\delta B^{(V)a\mu}}}\Big|_{B^{(V,A)}=0},\qquad
A_\mu^a={{\delta\Gamma(U,A^L,B^{(V)},B^{(A)})}
\over{\delta B^{(A)a\mu}}}\Big|_{B^{(V,A)}=0}.
\end{eqnarray}
We should mention that gauging the anomalous part of the action (6)
allows for a non-minimal term which vanishes as long as only strong
fields are present (cf. eq. 3.11 of ref.$^{11}$.). This new term gives
rise for a new coupling constant $d_1$ which, however, may be determined
from the decay width of $\omega\rightarrow\pi^0\gamma$.$^{12}$

\subsection{Electromagnetic properties of $\frac{1}{2}^+$ baryons}

Since the generator for the $U_{\rm e.m.}(1)$ gauge transformation
in SU(3) is given by $Q_{\rm e.m.}=diag(2/3,-1/3,-1/3)$ the
electromagnetic current $J_\mu^{\rm e.m.}$ is related to a special
linear combination of vector currents:
\begin{eqnarray}
J_\mu^{\rm e.m.}=V_\mu^3+{1\over{\sqrt3}}V_\mu^8.
\end{eqnarray}
The electromagnetic form factors of baryon $B$ are defined as the matrix
elements of $J_\mu^{\rm e.m.}$:
\begin{eqnarray}
{{\sqrt{p_0p_0^\prime}}\over{M_B}}\langle B({\bf p}^\prime)
|J_\mu^{em}(0)| B({\bf p})\rangle=
i{\overline u}({\bf p}^\prime)\big[\gamma_\mu F_1^B(Q^2)+
{{\sigma_{\mu\nu}Q^\nu}\over{2M_B}}F_2^B(Q^2)\big]u({\bf p})
\end{eqnarray}
where $Q_\mu=p_\mu-p_\mu^\prime$ is the momentum transfer and $M_B$ is the
physical mass of baryon $B$.  For the discussion of electromagnetic properties
the introduction of electric, $G_E^B(Q^2)$, and magnetic, $G_M^B(Q^2)$, form
factors is more suitable:
\begin{eqnarray}
&G_E^B(Q^2)&=F_1^B(Q^2)-{{Q^2}\over{4M_B^2}}F_2^B(Q^2)\cr
&G_M^B(Q^2)&=F_1^B(Q^2)+F_2^B(Q^2).
\end{eqnarray}
At zero momentum transfer the electric form factor yields the charge of $B$,
while the magnetic moment is obtained from:
\begin{eqnarray}
\mu_B=G_M^B(0).
\end{eqnarray}
To evaluate  $G_M^B(Q^2)$ we require the Fourier transformation of the space
components of the electromagnetic current, $J_i^{\rm e.m.}$ in the Breit frame.
We thus substitute the {\it ans\"atze} for the pseudoscalar and vector meson
fields into the expression for the current gained from eq. (29).
The angular velocities are eliminated via the quantization rule (24) and
matrix elements of the operators are evaluated using the {\it exact}
eigenstates of the collective Hamiltonian (26). Details of this
calculation as well as analytic expressions for the currents are
given in ref.$^{30}$ .

Table 2 contains our numerical results for the
magnetic moments using the parameters (28).
\begin{table}
\tcaption{The magnetic moments for the spin ${1\over2}$ baryons
predicted by our model compared to experimental numbers. Also the results
obtained in a model of pseudoscalars(PS) only are listed (ref.$^{20}$).}
\centerline{\tenrm\smalllineskip
\begin{tabular}{l c c c}\\
& This model & Expt. & PS \\
\hline
$p$ & 2.36 & 2.79 & 2.03 \\
$n$ & -1.87 & -1.91 & -1.58 \\
$\Lambda$  & -0.60 & -0.61 & -0.71 \\
$\Sigma^+$ & 2.41  & 2.42$\pm$0.05 & 1.99 \\
$\Sigma^0$ & 0.66  & ---        & 0.60 \\
$\Sigma^-$ & -1.10 & -1.16$\pm$0.03 & -0.79 \\
$\Xi^0$    & -1.96 & -1.25$\pm$0.01 & -1.55 \\
$\Xi^-$    & -0.84 & -0.68$\pm$0.03 & -0.64 \\
$\Sigma^0\rightarrow\Lambda$ &-1.74 & -1.61$\pm$0.08 & -1.39\\
\hline\\
\end{tabular}}
\end{table}
As expected, the incorporation of vector mesons supplies the desired
effect of increasing the contribution from the isovector current,
$V_i^3$ to the magnetic moments while leaving the isoscalar part,
$V_i^8$ almost unaltered. This leads to improved predictions for the
magnetic moments of $p,n,\Lambda,\Sigma^+,\Sigma^-$ and also the
$\Sigma^0\rightarrow\Lambda$ transition moment. However, we observe
that most of our predictions do not deviate significantly from
the SU(3) relations:$^{34}$
\begin{eqnarray}
&\mu_{\Sigma^+}&=\mu_{p}\quad \mu_{\Sigma^0}={1\over2}
(\mu_{\Sigma^+}+\mu_{\Sigma^-}),\quad \mu_{\Sigma^-}=\mu_{\Xi^-},\cr
&2\mu_\Lambda&=-(\mu_{\Sigma^+}+\mu_{\Sigma^-})
=-2\mu_{\Sigma^0}=\mu_n=\mu_{\Xi^0}=
{2\over{\sqrt3}}\mu_{\Sigma^0\rightarrow\Lambda}
\end{eqnarray}
although the baryon wave functions are strongly distorted SU(3)
$D$-functions. A major disagreement from the relations (34) is only
found for the relation involving
the neutron whose magnetic moment, in agreement with experiment, is much
smaller than $-(\mu_{\Sigma^+}+\mu_{\Sigma^-})$. Unfortunately
$\mu_n=\mu_{\Xi^0}$ is still satisfied approximately, predicting a too
large value for $\mu_{\Xi^0}$. Also the deviation from $\mu_p=
\mu_{\Sigma^+}$ is only small and opposite to the direction required by
experiment, resulting in a somewhat too small $\mu_p$. Recent investigations
have shown that a deviation from (34) in the desired direction and thus a
fine tuning of the magnetic moments may be achieved
when the shape of the profile functions is sensitive to symmetry
breaking.$^{35,36}$ Nevertheless, we may state that our
model provides a reasonable description of the magnetic moments of
${1\over2}^+$ baryons.

The vector meson fields play an even more pronounced role for the momentum
dependence of the form factors. This is, for example, indicated by the
magnetic radii:
\begin{eqnarray}
\langle r^2\rangle_M^B=-{6\over{\mu_B}}{{\partial G_M^B(Q^2)}
\over{\partial Q^2}}
\end{eqnarray}
the results for different baryons are listed in table 3.
\begin{table}
\tcaption{Same as table 2) for the magnetic radii defined in eq. 35.}
\centerline{\tenrm\smalllineskip
\begin{tabular}{l c c c}\\
& This model & Expt. & PS \\
\hline
$p$ & 0.94 & 0.74 & 0.43 \\
$n$ & 0.94 & 0.77 & 0.43 \\
$\Lambda$  & 0.78 & --- & 0.36 \\
$\Sigma^+$ & 0.96  & --- & 0.45 \\
$\Sigma^0$ & 0.86  & ---  & 0.36 \\
$\Sigma^-$ & 1.07 & --- & 0.58 \\
$\Xi^0$    & 0.90 & --- & 0.43 \\
$\Xi^-$    & 0.84 & --- & 0.38 \\
$\Sigma^0\rightarrow\Lambda$ &0.97 & --- & 0.48\\
\hline\\
\end{tabular}}
\end{table}
As shown in ref.$^{22}$ vector mesons provide an additional contribution to
the squared radii (of the order $6/m_V^2$) due to a mechanism comparable
to VMD. This effect also shows up in the SU(3) model thereby slightly
overestimating the experimental values for the magnetic radii of the
proton and the neutron.  Also the fact that we are dealing with a large
soliton is recognizable from the predicted values of the magnetic radii.
In any event the vector meson model gives significantly better agreement
with experiment than the model containing only pseudoscalar fields.

Next we consider the electric form factors $G_E^B(Q^2)$ which are
obtained as the Fourier transform of $J_0^{\rm e.m.}$. Again we
employ the exact eigenstates of the collective Hamiltonian to evaluate
the matrix elements of the relevant SU(3) operators.
In table 4 we display the results for the electric radii defined by:
\begin{eqnarray}
\langle r^2\rangle_E^B=-6{{\partial G_E^B(Q^2)}\over{\partial Q^2}}.
\end{eqnarray}
The absolute values of the radii of charged particles are drastically
increased by the inclusion of vector mesons demonstrating once again
that the soliton is quite large. On the other hand the experimental
value for the neutron electric radius is nicely reproduced.
\begin{table}
\tcaption{Same as table 2) for the electric radii defined in eq. 36.}
\centerline{\tenrm\smalllineskip
\begin{tabular}{l c c c}\\
& This model & Expt. & PS \\
\hline
$p$ & 1.20 & 0.74 & 0.59\\
$n$ & -0.15 & -0.12 & -0.22\\
$\Lambda$  & -0.06 & --- & -0.08\\
$\Sigma^+$ & 1.20  & --- & 0.59\\
$\Sigma^0$ & -0.01  & ---  & -0.02\\
$\Sigma^-$ & -1.21 & --- & -0.63\\
$\Xi^0$    & -0.10 & --- & -0.15\\
$\Xi^-$    & -1.21 & --- & -0.49\\
\hline\\
\end{tabular}}
\end{table}
This may also be observed from figure 4 where the momentum dependence of
the proton and neutron electic form factors are shown.
\begin{figure}[t]
\centerline{\hskip -1.5cm\psfig{figure=effpt.ps,height=9.0cm,width=16.0cm}}
\fcaption{The electric form factors of the neutron ($n$) and the
proton ($p$) as functions of the momentum transfer $Q$. The
solid lines correspond to the model's predictions. The dashed lines
are the dipole fit $1/(1+Q^2/0.71{\rm GeV}^2)^2$ for the proton
and the fit of Platchkov et al.$^{37}$ for the neutron. Also shown are some
selected expermental data for the neutron electric form factor. Note
that all numbers for the neutron electric form factor have been
scaled by a factor 10.}
\end{figure}
While there is
some deviation of the predicted proton electric form factor from the
dipole fit: $1/(1+Q^2/0.71{\rm GeV}^2)^2$, the neutron electric form
factor follows closely the phenomenological parametrization of
Platchkov et al.$^{37}$ who extracted the neutron electric form factor from
the deuteron form factor using Paris potential wave functions:
\begin{eqnarray}
G_E^n(Q^2)=-1.25\mu_n{{\tau}\over{1+18.3\tau}}{1\over{(1+Q^2/0.71GeV^2)^2}},
\ \tau={{Q^2}\over{4M_N^2}}
\end{eqnarray}
up to momentum transfers of about $0.3GeV$ possessing, however, a
somewhat smaller maximum than the Platchkov et al.$^{37}$ fit.
Summarizing this subsection on electromagnetic properties we may state
that the model provides reasonable predictions for the magnetic moments
and the corresponding radii. However, we suspect from the discussion of
the proton electric form factor that for charged baryons the absolute
values for the electric radii are overestimated.

\subsection{Strangeness in the proton}

Having seen that the vector meson model extended to SU(3) gives a fair
description of the static electromagnetic properties of $\frac{1}{2}^+$
baryons we shall now proceed and employ this model for predictions on
the relevance of strange quarks in the nucleon. {\it I.e.} we will
evaluate matrix elements of operators involving strangeness between
nucleon states.
Let us firstly consider the strange vector current:
\begin{eqnarray}
V_\mu^s=V_\mu^0-{2\over{\sqrt3}}V_\mu^8,
\end{eqnarray}
which in the quark language corresponds to ${\overline s}\gamma_\mu s$.
The matrix elements of $V_\mu^s$ are of special interest for
precision analyses of $eN$ and $\nu N$ scattering experiments$^{6}$
in the standard model. Again it proves useful to define
Dirac ($F_3(Q^2)$) and Pauli ($\tilde F_3(Q^2)$) type form factors
for the strangeness current $V_\mu^s$ which are the analogues of
$F_1^p(Q^2)$ and $F_2^p(Q^2)$ in eq. 31, respectively. Obviously we have
$F_3(0)=0$ since the net strangeness in the proton is zero. For
the corresponding radius which is defined as the slope of the
Dirac form factor:
\begin{eqnarray}
\langle r_s^2\rangle =
-6{{\partial F_3(Q^2)}\over{\partial Q^2}}\big|_{Q^2=0}
\end{eqnarray}
we find numerically $\langle r_s^2\rangle =0.05 fm^2$. Amusingly this
result has the opposite sign of the prediction in the pseudoscalar
model$^{20}$:$-0.10fm^2$. The positiveness of $\langle r_s^2\rangle$ may be
linked to VMD since a one pole VMD approach$^{20}$ together with
$\omega-\phi$ mixing leads to $\langle r_s^2\rangle=0.01fm^2$; which,
however, is much smaller in magnitude. Also Jaffe$^{38}$ obtained (by using
a three pole vector meson fit to dispersion relations suggested by
H\"ohler et al.$^{39}$) a positive value
$\langle r_s^2\rangle=0.16\pm0.06fm^2$,
although this is larger than our result. Actually the fit of ref.$^{38}$
has the unusual feature that the $\phi$ meson has a fairly substantial
coupling to the nucleon.

Our prediction for the strange magnetic moment of the proton:
\begin{eqnarray}
\mu_s=\tilde F_3(0)=-0.05
\end{eqnarray}
has the same sign as in the model of pseudoscalars only$^{20}$: $-0.13$ with a
smaller absolute value. Our result is even much smaller (in magnitude) than
Jaffe's estimate$^{38}$: $-0.31\pm0.09$.

We should stress once again that all these results are obtained by taking
matrix elements between exact eigenstates of the Hamiltonian (26) including
symmetry breaking. It is, of course, interesting to ask how symmetry
breaking influences the matrix elements of strange operators. Using the
SU(3) symmetric wave function of the proton yields $\mu_s=-0.54$,
{\it i.e.} the distorted wave function reduces the strange magnetic
moment by about $90\%$! On the other hand the already small result for
$\langle r_s^2\rangle$ is only mildly changed to $0.04fm^2$ when the
symmetric wave-function is employed.

Turning to the scalar strangeness content of the nucleon we are immediately
led to the puzzle of the nucleon $\Sigma$ term. The nucleon $\Sigma$
term is defined by:
{\begin{eqnarray}
\Sigma_\pi(Q^2)={1\over3}\sum_{i=1}^3\langle P({\bf p}^\prime)|
[Q^5_i,[Q^5_i,H]]|P({\bf p})\rangle
\end{eqnarray}
wherein $Q^5_i$ are the generators of an infinitesimal axial transformation
and $|P({\bf p})\rangle$ denots a proton state of momentum ${\bf p}$. Using
SU(3) symmetric wave functions for the baryons it has been observed
long ago$^{41}$ that
\begin{eqnarray}
\Sigma_\pi(0)=\frac{3\hat m}{\hat m-m_s}\big(M_\Lambda-M_\Xi\big)
\frac{1}{1-y},\quad {\rm with}\quad
y=\frac{2\langle {\overline s}s \rangle_P}
{\langle {\overline u}u + {\overline d}d \rangle_P}.
\end{eqnarray}
$y$ obviously measures the scalar strange content fraction in the proton.
Analyses of the $S$-wave amplitude in $\pi-N$ scattering leads to$^{42}$
$\Sigma_\pi(0)\approx45MeV$\footnote{The authors of ref.$^{42}$ pointed out
that further low-energy measurements would be needed to reduce the
uncertainty in $\Sigma_\pi(0)$.}. The puzzle now is obvious: Using the
experimental mass difference $M_\Xi-M_\Lambda=202MeV$ as well as
established values for the ratio ${\hat m}/m_s=25...30$ one finds
$y\approx0.4$ which, of course, is unexpectedly large. Skyrme
type models provide a solution to this puzzle by incorporating
SU(3) symmetry breaking effects. Working out the double commutator (41)
gives the generic form of the nucleon $\Sigma$-term:
\begin{eqnarray}
\Sigma_\pi(Q^2)=\frac{4}{3}I(Q^2)\Big(2+\langle D_{88}\rangle_P\Big)
\end{eqnarray}
wherein $I(Q^2)$ is the Fourier transform the integrand of the
symmetry breaker $\gamma$, especially we have
$I(0)=\frac{3\delta^\prime}{8(\delta^{\prime\prime}-\delta^\prime)}
\gamma$. Using eq. 9 and performing a perturbation expansion$^{18}$
of the matrix element $\langle D_{88}\rangle_P$ in terms of the
effective symmetry breaking parameter $\beta^2\gamma$ we find:
\begin{eqnarray}
\Sigma_\pi(0)=\frac{3\hat m}{\hat m-m_s}\big(M_\Lambda-M_\Xi
-0.031\gamma(\beta^2\gamma+\cdot\cdot\cdot)\big)
\end{eqnarray}
the dots refer to higher order terms in $\beta^2\gamma$. Values for
$\beta^2$ and $\gamma$ which fit the baryon mass spectrum reasonable
well are of the order $\beta^2\approx4GeV^{-1}$ and $\gamma\approx
1.5GeV$. This demonstrates that SU(3) symmetry breaking effects
may easily double the ``na\"\i ve" estimate (42) for the nucleon
$\Sigma$-term.

Of course, we shall carry out the exact evaluation of $\Sigma_\pi(0)$
which also takes into account the dots in eq. 44. Actually, our
prediction for $M_\Xi-M_\Lambda$ is somewhat larger than the
experimental value (cf. table 1).  Also, due to the large
soliton, our numerical results for $\beta^2$ and $\gamma$ are
slightly larger than the above mentioned estimates. This
results in an even increased value for $\Sigma_\pi(0)=62MeV$.
Recently Gasser et al.$^{43}$ have converted the difference between
the value at the Cheng-Dashen point, $\Sigma_\pi(t^2=
-{\bf Q}^2=2m_\pi^2)$\footnote{$\Sigma_\pi (-{\bf Q}^2=2m_\pi)$ has been
extracted from $\pi N$ scattering data by Koch$^{44}$ to be $64\pm8MeV$;
interpolating the $\pi N$ phase shifts from Weinberg's$^{45}$ result for
the scattering length $a_0^0=0.16m_\pi^{-1}$ reduces this result
to $56\pm8MeV$.}\,\, and zero momentum transfer into a radius (employing
constraints from chiral perturbation theory). They obtained the
surprisingly large value $\langle r^2_\Sigma\rangle=1.6fm^2$
which is about twice as large as the experimental value for
$r_E^2$ (cf. section 4). However, our result for
$\langle r^2_\Sigma\rangle=1.03fm^2$ is actually smaller than the
predicted electric radius of the proton (cf. table 3). Here it is,
of course, appropriate to compare the results of ref.$^{43}$ with our
predicted differences
$\Sigma_\pi(t^2=-{\bf Q}^2=2m_\pi^2)-\Sigma_\pi(0)=13.3 MeV\ (\approx15MeV)$
and $\Sigma_\pi(0)-\Sigma_\pi(t^2=-{\bf Q}^2=2m_\pi^2)=9.2MeV\ (\approx10MeV)$;
the numbers in parentheses are extracted from fig.3 of ref.$^{43}$. This
seems to indicate that our smaller prediction for
$\langle r^2_\Sigma\rangle$ is linked to the too large value for
$\Sigma_\pi(0)$ which enters in the denominator in the definition of
$\langle r^2_\Sigma\rangle$. We may speculate about the origin of the
large $\Sigma_\pi(0)$. In the calculation of $\Sigma_\pi$ we have
restricted ourselves to leading order in ${1\over{N_C}}$.
On the other hand the $\Sigma_\pi$-term is extracted from
the $s$-wave amplitude in $\pi N$ scattering which can only be
described reasonably well in soliton models when higher order
corrections in ${1\over N_C}$ are taken into account.$^{46}$ These
higher order corrections may thus also have a significant
influence on the $\Sigma$-term. Also, the inclusion of explicit scalar
degrees of freedom (like {\it e.g.} via the trace anomaly$^{47}$) might
alter the results since the $\Sigma$-term is associated to the matrix
element  $<p|{\overline u}u+{\overline d}d|p>$. Technically this is
realized by scaling the symmetry breaking terms in eq. 7 with powers
of ${\rm exp}(\sigma)<1$ wherein $\sigma$ is a gluonium scalar
order parameter which is introduced to mock up the QCD trace anomaly.
This procedure reduces the integrand of $I(Q^2)$ and thus also
$\Sigma_\pi$. Surely, this is only speculative since the incorporation
of scalar degrees of freedom is somewhat arbitrary and might change
other baryon observables as well.

Since we have seen that $\Sigma_\pi$ is related to the scalar
strange content fraction$^{48,49}$ $X_S$ of the proton we will also quote the
predictions of our model for $X_S$. Using a $\sigma$ - model interpretation
of the quark bilinears we define:
{\begin{eqnarray}
X_S={{<p|{\overline s}s|p>-<0|{\overline s}s|0>}\over
{<p|{\overline u}u+{\overline d}d+{\overline s}s|p>-
<0|{\overline u}u+{\overline d}d+{\overline s}s|0>}}=
{1\over3}\langle p|1-D_{88}|p\rangle.
\end{eqnarray}
Evaluating this matrix element with the exact eigenstates of the
collective Hamiltonian yields $X_S=0.134$. This is still fairly large
although it already represents an almost $50\%$ reduction compared to
the value $X_S={7\over{30}}\approx0.233$ which is obtained by using the
SU(3) symmetric proton wave function.$^{48}$ In the language of
eq. 42 the obtained value $X_S=0.134$ corresponds to $y=0.31$.

\subsection{Axial current matrix elements}

To discuss matrix elements of the axial current, $A_\mu^a$ (corresponding to
${\overline q}\gamma_\mu\gamma_5{{\lambda_a}\over2}q$ in the quark language)
between proton states we define the proton axial form factors of the
$l^{th}$ axial quark current
\begin{eqnarray}
{{\sqrt{p_0p_0^\prime}}\over{M_P}}\langle P({\bf p}^\prime)|
{\overline q}_l\gamma_\mu\gamma_5q_l|P({\bf p})\rangle =
{\overline u}({\bf p}^\prime)\big[\gamma_\mu\gamma_5H_l(Q^2)+
{{Q_\mu}\over{2M_P}}\gamma_5\tilde H_l(Q^2)\big]u({\bf p}),
\end{eqnarray}
with $l=1,2,3=u,d,s$ and $|P({\bf p})\rangle$ again denotes a proton state
with momentum ${\bf p}$. Here we will only be concerned with the main
form factor $H_l(Q^2)$ and therefore omit total derivatives in
the calculation of the axial current from eq. 29 since they only
contribute to the induced form factor $\tilde H_l(Q^2)$.$^{26}$ For the
evaluation of $H_l(Q^2)$ it is actually sufficient to only consider
the spatial components $A_i^a$. As in the case of the electromagnetic
properties we insert the {\it ans\"atze} made for the meson fields
eqs. 14 and 19-22 into the expression for axial current gained
from (29). Finally the Fourier transform of $A_i^a$ is
interpreted as the corresponding form factor in the Breit frame.

Of course, the final goal is to disentangle the three form factors
$H_l(Q^2)$. However, considering the linear combinations
\begin{eqnarray}
&G_a(Q^2)&=H_1(Q^2)-H_2(Q^2) \nonumber \\
&R(Q^2)&=H_1(Q^2)+H_2(Q^2)-2H_3(Q^2) \nonumber \\
&H(Q^2)&=H_1(Q^2)+H_2(Q^2)+H_3(Q^2)
\end{eqnarray}
turns out to be more instructive. The momentum dependence of these
form factors is displayed in figure 5.
\begin{figure}
\centerline{\hskip -1.5cm\psfig{figure=aaa.ps,height=9.0cm,width=16.0cm}}
\fcaption{The momentum dependence of the axial form factors
of the nucleon (cf. eq. 47). The input parameters are as in eq. 28.}
\end{figure}
The form factor for neutron $\beta$ decay, $G_a(Q^2)$ is obtained
by Fourier transforming $2A_3^3$. Its value at zero momentum
transfer $g_a=G_a(0)$ is very well known experimentally to be
$1.26$. There are no restrictive informations on $R(Q^2)$ and
$H(Q^2)$ from experiment. The only additional measurement on
axial from factors is the EMC experiment of scattering polarized
muons off polarized protons.$^{5}$ There the linear combination
$4H_1(Q^2)+H_2(Q^2)+H_3(Q^2)$ was investigated. Only by employing
SU(3) symmetry relations and informations on semileptonic
hyperon decays further inside into $R(Q^2)$, the Fourier
transform of $2\sqrt3A_3^8$, and $H(Q^2)$, the Fourier
transform of the axial singlet current $3A_3^0$, may be
gained. These investigations have led to the famous ``proton spin
crisis" stating the quarks' contribution to the proton spin, $2H(0)$,
might vanish identically.$^{50}$ However, we will abandon the use of
SU(3) symmetry relations and explore the effects of SU(3)
symmetry breaking in the baryon wave functions on these form factors
as well as for the Cabibbo model$^{51}$ which provides various relations
between the form factors of {\it strangeness changing} axial
currents.

Using the parameters listed in eq. 28 we find for the form factors at
zero momentum transfer:
\begin{eqnarray}
g_a=0.93,\ R(0)=0.38,\ H(0)=0.29.
\end{eqnarray}
Our result for $g_a$ is still about $30\%$ lower than the experimental value
although it represents a noticeable improvement against pseudoscalar
models.$^{20}$ A similar result for $g_a$ is already obtained in the
U(2) reduction of the present model indicating that strange degrees
of freedom only play a minor role for axial matrix elements of the
proton.$^{11}$
This gets more transparent by disentangling the individual quark form
factors at zero momentum transfer
\begin{eqnarray}
H_1(0)=0.63,\ H_2(0)=-0.31,\ H_3(0)=-0.03,
\end{eqnarray}
showing that strange quarks contribute as few as $10\%$ of the down quarks to
the matrix elements of the axial current.

Amusingly $g_A$ and $R(0)$ approximately satisfy the SU(3)
relation $R={3\over7}g_A(=0.40)$, which is obtained by comparing the
matrix elements of $D_{83}$ and $D_{33}$ assuming the wave functions
to be SU(3) symmetric. However, we already observe from figure 5
that the momentum dependence of $R$ follows more closely that of $H$ than
of $G_a$ and we will immediately show that this coincidence is
completely accidental for the parameter set (28). One striking
evidence for this behavior is observed form the radii of the
axial form factors:
\begin{eqnarray}
\langle r_3^2\rangle =0.51fm^2,\ \langle r_8^2\rangle =0.34fm^2,\
\langle r_0^2\rangle =0.35fm^2,
\end{eqnarray}
the lower index denotes the flavor component under consideration. The fact
that $\langle r_8^2\rangle$ is almost identical to
$\langle r_0^2\rangle$ indicates that the eighth and singlet components
of the axial current are both dominated by $H_1+H_2$. Note, that
on the contrary the above quoted SU(3) relation would enforce
$\langle r_3^2\rangle\approx\langle r_8^2\rangle$. The numerical
results (48) and (50) seem to justify the assumption
$R(Q^2)=H(Q^2)$ made in ref.$^{25}$ although there is no {\it a priori}
reason for this identification.  As in the
case of the electromagnetic radii the model gives a somewhat larger value
for the axial isovector radius $\langle r_3^2\rangle$ than the empirical
data$^{52}$: $0.40\pm0.04fm^2$.

We may now ask for the relevance of these predictions for the analysis
of the EMC experiment. In ref.$^{24}$ it was shown that $H(0)$ being as large
as $0.30$ is not in contradiction to the experimental data as long as
$R(0)$ is smaller than $0.5$; employing, however, the experimental value
for $g_a$. Let us therefore, for the time being, consider a different set
of parameters which gives the experimental value for $g_a$
\begin{eqnarray}
\tilde h=-0.1,\ \tilde g_{VV\phi}=1.4,\ \kappa=1.0
\end{eqnarray}
all other parameters remain as in (28). This set gives too large mass
differences, since we find $\gamma=2.545GeV$ yielding, {\it e.g.} $M_\Lambda-
M_N=236MeV$. Nevertheless (51) may shed some light on the analysis of the
EMC experiment. Although $g_A$ has increased to its experimental number, the
values for $R(0)$ and $H(0)$ have decreased to $0.29$ and $0.25$,
respectively. This shows once more that $R(0)$ is governed by the terms
it has in common with $H(0)$. The set (51) also leads to a very small
strange contribution $H_3(0)=-.015$ and a $50\%$ deviation from
the SU(3) relation between $g_a$ and $R(0)$. This impressively
demonstrates that employing SU(3) relations to relate matrix
elements of strangeness conserving currents to those of strangeness
changing currents in unreliable. We finally have to question whether
our treatment which obviously disagrees with one of the basic assumptions
underlying the interpretation of the EMC measurement is compatible with
the results of the EMC experiments. In our model we find
\begin{eqnarray}
{1\over9}\big[4H_1(0)+H_2(0)+H_3(0)\big]=\cases{
0.24\ {\rm for\ set\ eq.\ 28}\cr
0.28\ {\rm for\ set\ eq.\ 51}}
\end{eqnarray}
which actually is not very sensitive to the change of parameters and
compares reasonably well with the experimental value$^{5}$:$0.252\pm0.036$.

Finally let us also mention that additional information on $H_3(0)$ may be
extracted from elastic $\nu p$ scattering. Ahrens et al.$^{53}$ obtain $H_3(0)=
-0.15\pm0.09$, however, this result is claimed$^{54}$ to be subject
to change due to some theoretical uncertainties of analysis.

It remains to check whether or not this strong deviation from the SU(3)
relations contradicts the successful Cabibbo model for semileptonic
hyperon decays since in this model various strangeness changing currents
are related by assuming SU(3) symmetry. In such an approach only two
combinations (antisymmetric and symmetric) exist to construct flavor
singlets which are bilinear in the baryon octet and linear in the
octet currents. Establishing the electromagnetic charge normalization
completely fixes the structure of the vector current while for the
axial current two constants of proportionality, $F$ and $D$ for the
antisymmetric and symmetric combinations, respectively are left
undetermined. The relevant matrix elements at zero momentum transfer are
parametrized$^{55}$ as linear combinations of $F$ and $D$ (ignoring the spinor
bilinears $i{\overline u}({\bf p}^\prime)\gamma_\mu\gamma_5u({\bf p})$):
\begin{eqnarray}
&&\langle p|A^{k-}|\Lambda\rangle =-{{3F+D}\over{\sqrt6}},\
\langle \Sigma^-|A^{k-}|n\rangle =D-F\cr
&&\langle \Xi^-|A^{k-}|\Lambda\rangle ={{3F-D}\over{\sqrt6}},\
\langle \Xi^-|A^{k-}|\Sigma^0\rangle ={{F+D}\over{\sqrt2}},
\end{eqnarray}
furthermore relations for the diagonal matrix elements of the proton are
obtained: $g_a=F+D$ and $R(0)=3F-D$. The SU(3) symmetric Skyrme model
predicts$^{34}$ ${D/F}={9/5}$ which has been considered to be in nice
agreement with a typical fit$^{56,57}$ of (6.1) to experimental
data: ${D/F}= 1.63$. This agreement has been viewed as one of
the successes of the SU(3) symmetric Skyrme model. However, it has been
shown in several treatments$^{20,35,36,55}$ of the Skyrme model
of pseudoscalars that a reasonable description of hyperon decay
processes may even be achieved when the baryon wave functions deviate
strongly from SU(3).

Specifically we evaluate the matrix elements of $\int V_0^{k-}d^3r$ and
$\int A_i^{k-}d^3r$ between baryons, $B$ and $B^\prime$ of different
strangeness which yield the vector and axial vector charges
$g^-_v$ and $g^-_a$, respectively. Since $\int V_0^{k-}d^3r$
corresponds to the Noether charge, $g^-_v$ is just the matrix
element of the left SU(3) generator\footnote{The left
SU(3) generators are defined as $L_a=\sum_{b=1}^8D_{ab}R_b.$}
$L^-=L_4+iL_5$:
\begin{eqnarray}
g^-_v=\langle B^\prime|L^-|B\rangle.
\end{eqnarray}
The deviation of $g^-_v$ from its SU(3) symmetric value is
at least of second order in SU(3) symmetric breaking since
the matrix elements of $L_a$ vanish between states of different
SU(3) representations. This, of course, agrees with the
Ademollo-Gatto theorem.$^{58}$ Nevertheless, the deviations are not
necessarily small. Actually, experimental data do only exist for
the ratios $g^-_a/g^-_v$ and we present our results for the
two sets of parameters (28) and (51) in table 5 and compare the
predictions to those gained from the SU(3) relations.
\begin{table}
\tcaption{The $g^-_a/g^-_v$ ratios as predicted by our model. The
equation numbers refer to the input parameters. The SU(3) symmetry
predictions are obtained using $D/F=1.8$ as well as $F+D=g_a=1.26$.
The star indicates input quantities.}
\centerline{\tenrm\smalllineskip
\begin{tabular}{l c c c c}\\
&& $g^-_a/g^-_v$&& \\
&eq. 28 &eq. 51& Expt. & SU(3) \\
\hline
$\Lambda\rightarrow p$&0.69&0.74&0.70$\pm$0.03&0.74\\
$\Sigma^-\rightarrow n$&0.24&0.22&0.35$\pm$0.05&0.30\\
$\Xi^-\rightarrow\Lambda$&0.19&0.22&0.25$\pm$0.05&0.22\\
$\Xi^-\rightarrow\Sigma^0$&1.22&1.31&---&1.26\\
$n\rightarrow p$&0.93&1.26$^*$&1.26&1.26$^*$\\
$R$&0.38&0.29&---&0.66\\
\hline\\
\end{tabular}}
\end{table}
For the realistic set (28) the model describes the processes
$\Lambda\rightarrow p$ and $\Xi^-\rightarrow
\Lambda$ fairly well as the comparison with the data extracted from
experiment reveals. Our prediction of $g^-_A/g^-_V$ for the decay
$\Sigma^-\rightarrow n$ is on the low side which is mostly due to the
strong enhancement of $g^-_V$ for which the SU(3) symmetric wave
functions give unity (see also figure 1 of ref.$^{55}$). Changing to set (51)
which reproduces the experimental value of $g_a$
the $g^-_a/g^-_v$ ratios change only mildly. For the processes
$\Lambda\rightarrow p$ and $\Xi^-\rightarrow\Lambda$ we find again a fair
agreement with SU(3) predictions and experimental data. Unfortunately
$g^-_A/g^-_V$ deviates significantly from experimental data as well as
from SU(3) for $\Sigma^-\rightarrow n$ because $g^-_v=1.48$ is
almost $50\%$ larger than the SU(3) result while $g^-_a=0.327$ is
close to the SU(3) prediction ${2\over7}g_a=0.357$. However, we
should stress that this deviation from SU(3) is considerably smaller
than that found for $R(0)$.

Concluding this sub-section on axial form factors we may state that a
small value for the singlet matrix element of the proton may be explained
by Skyrme type models with a small polarization of the strange
sea of the proton once the condition of SU(3) symmetry is relaxed.
The deviations from SU(3) symmetry are sizable for strangeness
conserving matrix elements but small for those components of the
current which carry strangeness changing quantum numbers\footnote
{Similar results have also been found in ref.$^{59}$.}.
And therefore the extraction of $H_3(0)$ from the EMC data and SU(3)
symmetry relations is questionable.

\subsection{The neutron proton mass difference}

Finally let us consider isospin violation. In this context the
neutron proton mass difference is of special interest since it may be
used to get a deeper understanding of the problem discussed in the
preceding section, the ``proton spin puzzle". It has been demonstrated
that this mass difference may be employed to disentangle the ``matter"
and ``glue" contributions to the proton's axial singlet current matrix
element.$^{60}$ Here we shall not deal with this subject since it is covered
by another contribution to these proceedings.$^{61}$ Nevertheless, as a
first step it is appropriate to discuss the various contributions to this
quantities since their relative strengh is  crucial for the ``matter -glue"
decomposition of $H(0)$.

Experimentally the neutron proton mass difference is obtained to be
$1.29 MeV$. The strong as well as the electromagnetic interaction
contribute to this quantity:
\begin{eqnarray}
\big(M_{\rm neutron}-M_{\rm proton}\big)=\Delta_{\bf st.}+\Delta_{\bf e.m.}
\end{eqnarray}
Using experimental electromagnetic form factors $\Delta_{\bf e.m.}$
is estimated as$^{62}$: $(-0.76\pm0.30)MeV$. The negative sign may be
understood by noting that $\Delta_{\bf e.m.}$ is dominated by the Coulomb
repulsion. Thus we may extract:
\begin{eqnarray}
\Delta_{\bf st.}=(2.05\pm0.30)MeV.
\end{eqnarray}
which is the quantity we wish to explore in the vector meson model.
Actually it was found some time ago that vector mesons are
of utmost importance in order to predict a non-vanishing $\Delta_{\bf st.}$
in a two flavor Skyrme type model.$^{13}$ Of course, a two flavor model should
give the main share to $\Delta_{\rm st.}$ since one na\"\i vely expects
that $\Delta_{\bf st.}\sim m_d-m_u$.

We firstly have to incorporate isospin breaking in the meson sector before
discussing $\Delta_{\bf st.}$. This is done by generalizing the
symmetry breaking matrices in eq. 7:
\begin{eqnarray}
T+\frac{m_s}{\hat m}S\longrightarrow Y\lambda_3+T+\frac{m_s}{\hat m}S,
\end{eqnarray}
{\it i.e.} we have to introduce only one additional parameter
$Y=(m_u-m_d)/{\hat m}$ which may for example be fixed from the
$\omega-\rho^0$ mixing.$^{13,28}$

Having set up the isopin breaking part of the effective mesonic action
we return to the baryon sector. After substitution of the {\it ans\"atze}
(14), (19-22) into this action and canonical quantization we obtain
the collective Hamiltonian including isospin violating terms:
\begin{eqnarray}
H=H_{I=0}+H_{I=1}.
\end{eqnarray}
$H_{I=0}$ is identical to the Hamiltonian derived in section 3. The
isospin violating terms read:
\begin{eqnarray}
H_{I=1}&=&\Gamma_3D_{38}+
\Delta_3\sum_{i=1}^3(D_{3i}D_{8i}+D_{38}D_{88})
\nonumber \\
&&+{{\alpha_3}\over{\alpha^2}}\sum_{i=1}^3D_{3i}(R_i+\alpha_1D_{8i})
+{{\beta_3}\over{\beta^2}}\sum_{\alpha=4}^7
D_{3\alpha}(R_\alpha+\beta_1D_{8\alpha}).
\end{eqnarray}
Analytic expressions for the parameters measuring isospin breaking
in the baryon sector, $\Gamma_3,\Delta_3,\alpha_3$ and $\beta_3$ may
{\it e.g.} be traced in the appendix of ref.$^{28}$. $\Gamma_3$
and $\Delta_3$ are functionals
of the classical fields $F,\omega$ and $G$ while $\alpha_3$
and $\beta_3$ also contain the non-strange and strange excitations,
respectively. Note, however, that only the linear combination
$\eta_T\sim(\chi+\chi_8)$ appears in $\alpha_3$. In the evaluation
of the Hamiltonian only  terms linear
in the isospin breaking have been retained. Finally we have the
non-electromagnetic contribution to the neutron proton mass
difference:
\begin{eqnarray}
\Delta_{\bf st.}&=&\langle n|H_{I=1}|n \rangle
-\langle p|H_{I=1}|p \rangle
\nonumber \\
&=&-2\Gamma_3\langle p|D_{38}|p \rangle
-2\Delta_3\langle p|\sum_{i=1}^3(D_{3i}D_{8i}+D_{38}D_{88})|p \rangle
\nonumber \\
&&-2{{\alpha_3}\over{\alpha^2}}\sum_{i=1}^3
\langle p|D_{3i}(R_i+\alpha_1D_{8i})|p \rangle
-2{{\beta_3}\over{\beta^2}}\sum_{\alpha=4}^7
\langle p|D_{3\alpha}(R_\alpha+\beta_1D_{8\alpha})
|p \rangle,
\end{eqnarray}

Let us now discuss the numerical results for $\Delta_{\bf st.}$. For the
best fit parameters (28) we obtain:
\begin{eqnarray}
\Delta_{\bf st.}=1.65MeV
\end{eqnarray}
which is somewhat smaller than the value extracted from experimental
data (56). The contribution due to the classical fields in $\Gamma_3$
is only $0.20MeV$ since the matrix element $\langle p|D_{38}|p\rangle$
turns out to be reduced to about a quarter of its SU(3) symmetric value
$\sqrt3/30$. Of course, we use the exact proton eigenstate of $H_{I=0}$
to evaluate the matrix element of $H_{I=1}$. The dominant contribution
is due to the $\alpha_3$ term ($0.95MeV$)\footnote{In ref.$^{13}$ the
neutron proton mass difference in the U(2) reduction of the current
model was investigated. For the $\eta$ meson the {\it ad hoc} mass
of $549MeV$ was assumed.$^{12}$ This gives an even larger value: $1.10MeV$.}.
This is the only term which survives in the two flavor limit. In this
limit we have
$\sum_{i=1}^3\langle p|D_{3i}(R_i+\alpha_1D_{8i})|p\rangle=
\langle p|T_3|p\rangle=\frac{1}{2}$.
Its crucial to note that the $\eta$ mesons give the main share to this
part. Setting $\chi\equiv\chi_8\equiv0$ in eq. 19 and extremizing the
non-strange moment of inertia with this {\it ansatz} only gives $0.38MeV$!
This value is now solely due to vector mesons.
It is actually the fact that $\eta_T$ needs to be increased by about
$60\%$ in order to reproduce the experimental value for $\Delta_{\bf st.}$
which sheds some light on the possible decomposition of the singlet axial
form factor of the proton into ``matter" and ``glue" portions.$^{60,61}$
Finally let us mention that the strange excitations (through $\beta_3$)
contribute about $0.50MeV$ to the strong interaction part of the
neutron proton mass difference.

\section{Conclusions}

We have presented the treatment of the SU(3) Skyrme model including
vector mesons in the context of the collective approach which considers
SU(3) symmetry as approximative. This assumption motivates the introduction
of collective coordinates for the whole SU(3) group although these
coordinates no longer correspond to real zero modes.
Special care has been taken of fields which vanish at the classical
level but get excited by the collective rotation of the static
soliton. SU(3) symmetry breaking has been treated exactly on the
level of the collective Hamiltonian by employing a generalized
Yabu-Ando method for diagonalization. The resulting eigenstates
are no longer pure octet (decouplet) states but contain significant
admixtures of higher dimensional representations like ${\overline{10}}$
or $27$ $(27,35,{\overline{35}})$. The experimental numbers
for the mass differences of the low lying ${1\over2}^+$ and ${3\over2}^+$
baryons have been reproduced within $10\%$ although a quite large
soliton was needed to achieve that goal.

Vector and axial vector currents have been constructed by gauging the
mesonic action with external fields. Matrix elements of the
currents have yielded predictions for the static properties of the baryons.
Especially good agreement with experiment has been obtained for the
magnetic moments. Amusingly the results do not deviate strongly from
the predictions obtained by assuming SU(3) symmetry. Nevertheless
the strange magnetic moment of the proton has come out very small,
indicating, as na\"\i vely expected,
that strange effects in the nucleon are strongly suppressed.
Also reasonable results for the magnetic radii were obtained while the
the electric radii of charged spin $1/2$ baryons seemed to be
somewhat overestimated. This effect is most likely due to the
large soliton needed to fit the mass difference. Exposing the effects
of symmetry breaking in the context of matrix elements of the axial
currents was somewhat tricky since for the standard set of parameters (28)
the SU(3) relation between $g_a$ and $R(0)$ has accidentally been
recovered. Choosing, however, to work with a set of parameters which
gives the experimental value for $g_a$ reveals that the form factor
$R(Q^2)$ is closely related to the form factor of the axial singlet
current, $H(Q^2)$. In this way the model predicts a very small
contribution of strange quarks to the matrix element of the axial
current between nucleon states contradicting neither the EMC experiment
nor the successful Cabibbo model of semileptonic hyperon decays.

Let us finally comment on new developments concerning symmetry
breaking in the meson sector. In the current model the predicted
ratio $2m_s/{\hat m}\approx36$ is rather large compared to {\it e.g.}
the chiral perturbation program of Gasser and Leutwyler$^{62}$ who find
$25.0\pm2.5$. The small value of ref.$^{62}$ is mostly due to
the additional term
\begin{eqnarray}
&&Tr\Big[(\lambda^\prime T +\lambda^{\prime\prime} S)U
(\lambda^\prime T +\lambda^{\prime\prime} S)U\nonumber \\
&&\qquad+(\lambda^\prime T +\lambda^{\prime\prime} S)U^\dagger
(\lambda^\prime T +\lambda^{\prime\prime} S)U^\dagger-
2(\lambda^\prime T +\lambda^{\prime\prime} S)^2\Big]
\end{eqnarray}
which is of second order in SU(3) symmetry breaking. However this
term has only a very minor effect on the baryon mass spectrum
in the soliton picture. More crucial is the incorporation
of a symmetry breaking term for the wave-functions of the
vector mesons:
\begin{eqnarray}
Tr\Big[(\gamma^\prime T^+ +\gamma^{\prime\prime} S^+)
F_{\mu\nu}(\rho)F^{\mu\nu}(\rho))\Big]
\end{eqnarray}
which is needed to give the correct ratio $\Gamma(K^*)/\Gamma(\rho)$.
We have performed first calculations with this term in the soliton
sector.$^{28}$
\begin{table}
\tcaption{Best fit to the experimental baryon mass differences
with the symmetry breaking term (63) included.}
\centerline{
\begin{tabular}{l c c c c c c c }
&$\Lambda$&$\Sigma$&$\Xi$&$\Delta$&$\Sigma^*$&$\Xi^*$&$\Omega$  \\
\hline
Fit&168&263&404&327&470&617&766 \\
Expt.&177&254&379&293&446&591&733 \\
\hline
\end{tabular}}
\end{table}
In table 6 we present the corresponding best fit to the mass
spectrum which is gained for parameters
\begin{eqnarray}
\tilde h&=&0.36,\quad \tilde g_{VV\phi}=1.88,\quad \kappa=1.0
\end{eqnarray}
which are closer to the central values $\tilde h=0.4$ and
$\tilde g_{VV\phi}=1.9$ than eq. 28. Changing from set (28) to
(64) also decreases the size of the soliton somewhat. This is
seen most easily by noting that the baryonic radius is
reduced from $0.30fm^2$ to $0.28fm^2$. Thus we may speculate that
the inclusion of (63) improves the predictions for the baryon
radii. However, $g_a$ might come out even smaller than $0.93$.

The generalization of (63) to also include isospin breaking
increases the effects of vector mesons on the neutron proton
mass difference by about $30\%$ yielding $\Delta_{\rm st.}=1.77MeV$
which is within the experimental error bars (56).

\nonumsection{Acknowledgements}

The author would like to thank Prof. G. Holzwarth for
the warm hospitality extended at the workshop. Also the nice atmosphere
which stimulated a lot of fruitful discussions is appreciated.

The author furthermore wants to thank his colleagues J. Schechter
and N. W. Park who participated in parts of the work underlying this
talk and R. Alkofer for valuable remarks on the manuscript.

This work is supported by the Deutsche Forschungsgemeinschaft (DFG) under
contract Re 856/2-1.

\vfil\eject

\nonumsection{References}

\end{document}